# PyAutoFit: A Classy Probabilistic Programming Language for Model Composition and Fitting

**James. W. Nightingale**[1], **Richard G. Hayes**[1], **and Matthew Griffiths**[2]

**1** Institute for Computational Cosmology, Stockton Rd, Durham, United Kingdom, DH1 3LE **2** ConcR Ltd, London, UK

## Summary

A major trend in academia and data science is the rapid adoption of Bayesian statistics for data analysis and modeling, leading to the development of probabilistic programming languages (PPL). A PPL provides a framework that allows users to easily specify a probabilistic model and perform inference automatically. `PyAutoFit` is a Python-based PPL which interfaces with all aspects of the modeling (e.g., the model, data, fitting procedure, visualization, results) and therefore provides complete management of every aspect of modeling. This includes composing high-dimensionality models from individual model components, customizing the fitting procedure and performing data augmentation before a model-fit. Advanced features include database tools for analysing large suites of modeling results and exploiting domain-specific knowledge of a problem via non-linear search chaining. Accompanying `PyAutoFit` is the autofit workspace, which includes example scripts and the `HowToFit` lecture series which introduces non-experts to model-fitting and provides a guide on how to begin a project using `PyAutoFit`. Readers can try `PyAutoFit` right now by going to the introduction Jupyter notebook on Binder or checkout our readthedocs for a complete overview of **PyAutoFit**'s features.

## Background of Probabilistic Programming

Probabilistic programming languages (PPLs) have enabled contemporary statistical inference techniques to be applied to a diverse range of problems across academia and industry. Packages such as PyMC3 (Salvatier et al., 2016), Pyro (Bingham et al., 2019) and STAN (Carpenter et al., 2017) offer general-purpose frameworks where users can specify a generative model and fit it to data using a variety of non-linear fitting techniques. Each package is specialized to problems of a certain nature, with many focused on problems like generalized linear modeling or determining the distribution(s) from which the data was drawn. For these problems the model is typically composed of the equations and distributions that are fitted to the data, which are easily expressed syntactically such that the PPL API offers an expressive way to define the model and extensions can be implemented in an intuitive and straightforward way.

## Statement of Need

`PyAutoFit` is a PPL whose core design is providing a direct interface with the model, data, fitting procedure and results, allowing it to provide comprehensive management of many different aspects of model-fitting. **PyAutoFit** began as an Astronomy project for fitting large imaging datasets of galaxies, after the developers found that existing PPLs were not suited to the type of model fitting problems Astronomers faced. This includes efficiently analysing large and homogenous datasets with an identical model fitting procedure, making it straight forward to fit many models to large datasets with streamlined model comparison and massively parallel support for problems where an expensive likelihood function means run-times can be of order days or longer. More recent development has generalized `PyAutoFit`, making it suitable to a broader range of model-fitting problems.

## Software Description

To compose a model with `PyAutoFit` model components are written as Python classes, allowing `PyAutoFit` to define the model and associated parameters in an expressive way that is tied to the modeling software's API. A model fit then requires that a `PyAutoFit` `Analysis` class is written, which combines the data, model and likelihood function and defines how the model-fit is performed using a `NonLinearSearch`. The `NonLinearSearch` procedure is defined using an external inference library such as `dynesty` (Speagle, 2020), emcee (Foreman-Mackey et al., 2013) or `PySwarms` (Miranda, 2018).

The `Analysis` class provides a model specific interface between `PyAutoFit` and the modeling software, allowing it to handle the 'heavy lifting' that comes with writing model-fitting software. This includes interfacing with the non-linear search, outputting results in a structured path format and model-specific visualization during and after the non-linear search. Results are output in a database structure that allows the `Aggregator` tool to load results post-analysis via a Python script or Jupyter notebook. This includes methods for summarizing the results of every fit, filtering results to inspect subsets of model fits and visualizing results. Results are loaded as `Python` generators, ensuring the `Aggregator` can be used to interpret large files in a memory efficient way. `PyAutoFit` is therefore suited to 'big data' problems where independent fits to large homogeneous data-sets using an identical model-fitting procedure are performed.

## Model Abstraction and Composition

For many modeling problems the model comprises abstract model components representing objects or processes in a physical system. For example, galaxy morphology studies in astrophysics where model components represent the light profile of stars (Häußler et al., 2013; Nightingale et al., 2019). For these problems the likelihood function is typically a sequence of numerical processes (e.g., convolutions, Fourier transforms, linear algebra) and extensions to the model often requires the addition of new model components in a way that is non-trivially included in the fitting process and likelihood function. Existing PPLs have tools for these problems, for example 'black-box' likelihood functions in PyMC3. However, these solutions decouple model composition from the data and fitting procedure, making the model less expressive, restricting model customization and reducing flexibility in how the model-fit is performed.

By writing model components as Python classes, the model and its associated parameters are defined in an expressive way that is tied to the modeling software's API. Model composition with `PyAutoFit` allows complex models to be built from these individual components, abstracting the details of how they change model-fitting procedure from the user. Models can be fully customized, allowing adjustment of individual parameter priors, the fixing or coupling of parameters between model components and removing regions of parameter space via parameter assertions. Adding new model components to a `PyAutoFit` project is straightforward, whereby adding a new Python class means it works within the entire modeling framework.

`PyAutoFit` is therefore ideal for problems where there is a desire to compose, fit and compare many similar (but slightly different) models to a single dataset, with the `Aggregator` including tools to facilitate this.

For many model fitting problems, domain specific knowledge of the model can be exploited to speed up the non-linear search and ensure it locates the global maximum likelihood solution. For example, initial fits can be performed using simplified model parameterizations, augmented datasets and faster non-linear fitting techniques. Through experience users may know that certain model components share minimal covariance, meaning that separate fits to each model component (in parameter spaces of reduced dimensionality) can be performed before fitting them simultaneously. The results of these simplified fits can then be used to initialize fits using a higher dimensionality model. Breaking down a model-fit in this way uses `PyAutoFit`'s non-linear search chaining, which granularizes the non-linear fitting procedure into a series of linked non-linear searches. Initial model-fits are followed by fits that gradually increase the model complexity, using the information gained throughout the pipeline to guide each `NonLinearSearch` and thus enable accurate fitting of models of arbitrary complexity.

## History

`PyAutoFit` is a generalization of [PyAutoLens](), an Astronomy package developed to analyse images of gravitationally lensed galaxies. Modeling gravitational lenses historically requires large amounts of human time and supervision, an approach which does not scale to the incoming samples of 100000 objects. Domain exploitation enabled full automation of the lens modeling procedure ([Nightingale et al., 2018](); [Nightingale & Dye, 2015]()), with model customization and the aggregator enabling one to fit large datasets with many different models. More recently, `PyAutoFit` has been applied to calibrating radiation damage to charge coupled imaging devices and a model of cancer tumour growth.

## Workspace and HowToFit Tutorials

`PyAutoFit` is distributed with the [autofit workspace](), which contains example scripts for composing a model, performing a fit, using the `Aggregator` and `PyAutoFit`'s advanced statistical inference methods. Also included are the `HowToFit` tutorials, a series of Jupyter notebooks aimed at non-experts, introducing them to model-fitting and Bayesian inference. They teach users how to write model-components and `Analysis` classes in `PyAutoFit`, use these to fit a dataset and interpret the model-fitting results. The lectures are available on our [Binder]() and may therefore be taken without a local `PyAutoFit` installation.

## Software Citations

`PyAutoFit` is written in Python 3.6+ ([Van Rossum & Drake, 2009]()) and uses the following software packages:

- `corner.py` https://github.com/dfm/corner.py ([Foreman-Mackey, 2016]())
- `dynesty` https://github.com/joshspeagle/dynesty ([Speagle, 2020]())
- `emcee` https://github.com/dfm/emcee ([Foreman-Mackey et al., 2013]())
- `matplotlib` https://github.com/matplotlib/matplotlib ([Hunter, 2007]())
- `NumPy` https://github.com/numpy/numpy ([Harris et al., 2020]())
- `PyMulitNest` https://github.com/JohannesBuchner/PyMultiNest ([Feroz et al., 2009]()) ([Buchner et al., 2014]())

- `PySwarms` https://github.com/ljvmiranda921/pyswarms (Miranda, 2018)
- `Scipy` https://github.com/scipy/scipy (Virtanen et al., 2020)

## Related Probabilistic Programming Languages

- `PyMC3` https://github.com/pymc-devs/pymc3 (Salvatier et al., 2016)
- `Pyro` https://github.com/pyro-ppl/pyro (Bingham et al., 2019)
- `STAN` https://github.com/stan-dev/stan (Carpenter et al., 2017)
- `TensorFlow Probability` https://github.com/tensorflow/probability (Dillon et al., 2017)
- `uravu` https://github.com/arm61/uravu (McCluskey & Snow, 2020)

## Acknowledgements

JWN and RJM are supported by the UK Space Agency, through grant ST/V001582/1, and by InnovateUK through grant TS/V002856/1. RGH is supported by STFC Opportunities grant ST/T002565/1. This work used the DiRAC@Durham facility managed by the Institute for Computational Cosmology on behalf of the STFC DiRAC HPC Facility (www.dirac.ac.uk). The equipment was funded by BEIS capital funding via STFC capital grants ST/K00042X/1, ST/P002293/1, ST/R002371/1 and ST/S002502/1, Durham University and STFC operations grant ST/R000832/1. DiRAC is part of the National e-Infrastructure.